
\input phyzzx
\FRONTPAGE
\line{\hfill BROWN-HET-950}
\line{\hfill DAMTP 94-46}
\line{\hfill June 1994}
\vskip1.0truein
\titlestyle{{{\bf ELECTROWEAK BARYOGENESIS WITH TOPOLOGICAL DEFECTS}}
\foot{To appear in J.C. Rom\~{a}o (ed.), {\it Electroweak Physics and the Early
Universe}; Talk given by A.C. Davis at the NATO Advanced Research Workshop,
Sintra, Portugal, 23-25 March, 1994.}}
\bigskip
\author {Anne-Christine Davis\foot{e-mail:
A.C.Davis@damtp.cambridge.ac.uk}$^{1)}$, Robert Brandenberger\foot{e-mail:
rhb@het.brown.edu}$^{2)}$, Mark
Trodden\foot{e-mail: mtrodden@het.brown.edu}$^{2)}$ }
\vskip .25in
\item {1)} {\it Department of Applied Mathematics and Theoretical
Physics and Kings College, University of Cambridge, Cambridge CB3 9EW,
U.K.}
\item {2)} {\it Physics Department, Brown University, Providence, RI
02912, USA}

\endpage

{\bf \chapter{Introduction}}

Electroweak baryogenesis has become a topic of much recent activity
[1].  Here we discuss a new scenario which has the advantage of being
insensitive to the order of the electroweak phase transition. We briefly
review a mechanism [2] using unstable electroweak strings [3] and
discuss in detail a mechanism [4] using topological defects (in
particular cosmic strings) left behind after a previous phase
transition.

In the standard electroweak theory all three necessary conditions to
generate a net baryon number are satisfied.  Sphaleron transitions
violate baryon number [5].  The electroweak theory explicitly violates
C invariance and in extensions of the standard model with non-minimal
Higgs structure there is explicit CP violation [6].  Finally, out of
equilibrium field configurations may result as remnants of the phase
transition.

The key issue is how the `out of equilibrium' condition is realised.
In most previous work [7,8] use was made of bubble walls which form if
the electroweak phase transition is first order.  However, at present
it is unclear [9] whether the electroweak phase transition is
sufficiently strongly first order for baryogenesis mechanisms
involving bubble walls to be effective.  In [2] it was pointed out that
topological or non-topological defects may play a role similar to
bubble walls in triggering electroweak baryogenesis.  Such a mechanism
is independent of the order of the phase transition.

In the next section we show how electroweak baryogenesis can be
implemented in models with a second order phase transition and compare
it with the first order case.  A mechanism with metastable electroweak
strings is then briefly reviewed.  A more robust mechanism
using topological defects produced in a previous phase transition is
discussed and the resulting baryon asymmetry estimated.
Finally, we discuss future extensions to our work and conclusions.

{\bf \chapter{ELECTROWEAK BARYOGENESIS WITH A SECOND ORDER PHASE TRANSITION}}

Let us review how Sakharov's conditions are realised in electroweak
baryogenesis scenarios and compare the implementations in first and
second order phase transitions.

Baryon number violation occurs via sphaleron transitions.  The
transition rate is exponentially suppressed in the broken phase.
However, in the symmetric phase transitions are copious.  Their rate
per unit volume is [10]

$$\Gamma_B \sim \alpha_W^4 T^4\eqno {(2.1)}$$

\noindent where $\alpha_W = g^2/4\pi$, $g$ being the $SU(2)$ gauge coupling
constant.  In extensions of the standard electroweak theory with
non-minimal Higgs structure containing explicit $CP$ violation, there exists a
$CP$ violating phase which changes by an amount $\Delta\theta$ during the
phase transition.

In Fig.1 we compare the ways in which the out of equilibrium condition
is realized in models with first and second order phase transitions.
The key role is played by expanding bubble walls and contracting
topological defects respectively.  In the scenarios of Refs.1,2,
baryogenesis takes place in the outer edge of the bubble wall, ie\. where

$$\vert\phi \vert < g\,\eta_{EW}\eqno {(2.2)}$$

\noindent $\vert\phi\vert$ being the order parameter of the
transition.  The amplitude $\vert\phi\vert$ is increasing at any point
in space which the bubble wall crosses. This may be related to the change of
the $CP$-odd phase and hence $CP$ violation has a
preferred sign.  Finally, as long as the bubble wall moves at relativistic
speed, there will be no time to establish thermal equilibrium inside
the walls.

With a second order phase transition, the role of the bubble wall is
played by topological defects.  Baryogenesis takes place inside
the core of the defect and the surrounding region where (2.2) is
satisfied.  If the defects contract (and eventually evaporate), then
there will be an overall increase in $\vert\phi\vert$ and hence net
baryon number generation.  The field configurations within contracting
topological defects are out of thermal equilibrium.

In the next section we consider a specific implementation of this
using electroweak strings.

{\bf \chapter{BARYOGENESIS WITH ELECTROWEAK STRINGS}}

In [2] it was suggested that electroweak strings could provide the out
of equilibrium condition necessary to generate a baryon asymmetry.
The electroweak theory doesn't admit topologically stable strings.
However, it has been shown that it is possible to embed the
Nielsen-Olesen vortex [11] in the electroweak theory [12].  Such
strings could be metastable, in which case their formation would be
similar to that of topological strings.  The core of the string would
play a similar role to the bubble wall in first order phase
transitions.  In the core of the string the electroweak symmetry is
restored and baryon violating processes are unsuppressed with rate as
in (2.1).  Outside the string the electroweak symmetry is broken and
baryon violating transitions are exponentially suppressed.  Since the
strings are at best metastable and of finite length, terminating on a
monopole anti-monopole pair, they will collapse along their axis.
This collapse of the string provides the out of thermal equilibrium
condition.  If this is the case in the 2-doublet model then in the tip
of the string the $CP$ violating phase is rapidly changing.  Hence, we
have all the necessary conditions to generate a baryon asymmetry.

An optimistic estimate of the baryon asymmetry can be obtained by
assuming that the mean length and average separation of electroweak
strings at $t_G$, the time corresponding to the Ginsburg temperature
of the phase transition, is the correlation length, $\xi (t_G) \sim
\lambda^{-1} \eta^{-1}$.  The strings will collapse along their axes and
decay in time interval

$$\Delta t_s \sim v^{-1}(\lambda \eta)^{-1}\eqno {(3.1)}$$

\noindent where $v$ is the velocity of collapse, taken to be $\sim$ 1.
By considering the rate of change of the volume in which $CP$ violation is
effective
we can estimate the rate of baryon number generation per string to be

$${dN_B \over dt} \sim w^2 v \Gamma_B \Delta \theta \Delta t_c
\eqno {(3.2)}$$

\noindent where $w \sim \lambda^{-{1 \over 2}}\eta^{-1}$ is the string
width and $\Delta t_c$ is the length of time a fixed point in space is in the
transition
region.  Taking one string per correlation volume $\xi (t_G)^3$ and
integrating from the Ginsburg time $t_G$ to $t_G + \Delta t_s$ gives

$${n_B \over s} \sim {9 \over 2\pi^2g^*} \ {\lambda \over \gamma(v) v}
\Delta \theta g^3 \alpha^4_{W}\eqno {(3.3)}$$

\noindent where $g^*$ is the effective number of degrees of freedom.
In our estimate we have included a suppression for the fact that
baryon violating processes are only unsuppressed for $|\phi | < g\eta$.

For our mechanism to work we require the core radius to be large
enough to support sphaleron processes.  In addition, sphaleron
transitions must be suppressed in the broken phase for $T=T_G$.
Finally, we require the electroweak string to be metastable in the
2-doublet model.  Unfortunately, it has been shown [13] that this is
not the case for physical Weinberg angle [14].  Attempts to stabilise
it with quark and lepton condensates have been unsuccessful [15].
Hence, a more robust method of electroweak baryogenesis could involve
topological defects formed at a phase transition at a scale above the
electroweak scale.

{\bf \chapter{BARYOGENESIS WITH TOPOLOGICAL DEFECTS}}

In order to obtain topological defects -- cosmic strings to be
specific -- we assume that at an energy scale $\eta$, larger than
$\eta_{EW}$, there is another symmetry breaking which produces strings.
One possibility is to embed $SU(2) \times U(1)$ in some larger simply
connected group $G$ such that at a scale $\eta$

$$G\longrightarrow
SU(2)
\times
U(1)$$

\noindent and

$$\Pi_1
\left(
G/
(SU(2)
\times
U(1)
\right)
\not=
{\bf
1}$$

\noindent A second possibility is to assume that electroweak symmetry
breaking is induced dynamically by having a technifermion condensate
form at the scale $\eta_{EW}$:

$$ \langle \bar{\psi}_{TC} \psi_{TC} \rangle \not= 0 \, , \,\,\,\,
T\leq \eta_{EW} \, . \eqno$$

\noindent Here, $\psi_{TC}$ denotes the technifermion.  In this case
we can assume that fermion masses are induced by a second phase
transition in the technifermion sector at a scale $\eta$ which in
general is only slightly higher than $\eta_{EW}$.  It is possible that
strings form in this transition.  In the core of these strings the
fermion condensates vanish, the electroweak symmetry is unbroken, and
hence baryon number violating processes are unsuppressed.

The region of electroweak symmetry restoration is actually larger than
the core of the string.  In [16] it was shown that cosmic strings
formed at a previous phase transition and coupled to the Weinberg-Salam
model restore the electroweak symmetry out to a region of order
$\eta^{-1}_{EW}$. This is a result of the coupling between the string
fields and the electroweak gauge fields.  If the string is
superconducting then the region of symmetry restoration is much larger,
being given by [17,16].

$$R_s \sim I/\eta^2_{EW} \eqno {(4.1)}$$

\noindent where I is the current carried by the string.  The maximal
current is of order $\eta$.  (For grand unified strings this region is
macroscopic, being of order 10$^{-5}$ m!)

Let us now give a rough estimate of the baryon to entropy ratio which
can be generated using the proposed mechanism and compare the
result with that obtained by the mechanisms of Refs.7,8
which rely on bubble wall expansion. To simplify the calculations we
assume that the phase transition is rapid and that the strings move
relativistically (in order that the out-of-equilibrium condition is
satisfied).

The important parameters in our calculation are the total volume $V$,
the volume $V_{BG}$ in which net baryon number violating processes are
taking place, the rate $\Gamma_B$ of these processes (see (2.1)), and
the net change

$$\Delta
\theta
=
\int
dt
\dot{\theta}$$

\noindent in the $CP$ violating phase $\theta$. We are making the
plausible assumption that the electroweak symmetry is restored inside
the core of the string. In this case, the mean value of the CP
violating phase vanishes in the core. In the broken phase, the
distinguished value of $\theta$ will be nonvanishing. Hence, for
points in space initially inside the string core, the net change in
$\theta$ will have a preferred direction. In this respect there is no
difference between our mechanism and the ones of Refs.7,8.

The net baryon number density $\Delta n_B$ is then given by

$$\Delta n_B = {1\over V} \, {\Gamma_B\over T} \, V_{BG} \Delta \theta
\, .\eqno {(4.2)}$$

The volume $V_{BG}$ is determined by the mean separation $\xi$ of the
strings and the radius $R_s$ of the string core (strictly speaking the
part of the core where (2.2) is satisfied and hence $n_B$ violating
processes are unsuppressed).

The key to the calculation is a good estimate of $V_{BG}$.  Note that
the translational motion of a topological defect does not lead to any
net baryogenesis since $\Delta\theta = 0$ integrated over time.  At
the leading edge of the moving defect, a baryon number with one sign
will be produced, but at the trailing edge baryogenesis will have the
opposite sign.  We will return to this point later.  Contraction, on
the other hand, does produce a net $\Delta n_B$.  Integrated over
time, there is a net $\Delta\theta\not= 0$ in the entire volume
corresponding to the initial defect configuration. Hence, a lower
bound on the volume $V_{BG}$ is obtained by taking the volume occupied
initially by the collapsing defects at the time when baryogenesis
commences (see below).  We focus on string loops.  Their mean
separation is $\xi(t)$.  Hence, in one horizon volume

$$V = {4\pi\over 3} \, t^3 \, \eqno {(4.3)}$$

\noindent the corresponding volume where net baryon number generation
takes place is

$$V_{BG} \sim R_s^2 \xi (t) \left( {t\over \xi(t)} \right)^3 \,
.\eqno {(4.4)}$$

\noindent The last factor on the right hand side is the number of
string loops per horizon volume, the second factor is the length of a
loop.

Most of the contribution to the baryon to entropy ratio is generated
at a time $t_U$ soon after $t_{EW}$ when sphaleron processes cease to
be thermally excited in the true vacuum. To simplify the equations, we
will set the two times equal in the following.  Thus, to obtain an
order of magnitude estimate of the strength of our baryogenesis
mechanism we will evaluate all quantities at $t_{EW}$.  Combining
(4.2)-(4.4) and (2.1) yields

$$\Delta n_B (t_{EW} ) \sim \alpha_W^4 \Delta \theta \, \left(
{R_s\over \xi(t_{EW})}\right)^2 \, T_{EW}^3 \eqno {(4.5)}$$

\noindent or

$${\Delta n_B\over s} \sim {g^*}^{-1} \alpha_W^4 \Delta \theta \left(
{R_s\over \xi (t_{EW})} \right)^2 \equiv {g^*}^{-1} \alpha_W^4 \Delta
\theta \, (SF) \, ,\eqno {(4.6)}$$

\noindent with $g^*$ being the number of spin degrees of freedom in
radiation, and with

$$(SF) = \left( {R_s\over \xi (t_{EW})} \right)^2 \, .\eqno {(4.7)}$$

\noindent Apart from the factor $(SF)$, this is the same order of
magnitude as obtained in the mechanisms using a first order phase
transition [7,8].  Hence, we call $(SF)$ the ``suppression
factor".

Above, we implicitly assumed that all strings have the same radius.
This is a good approximation for strings in the friction dominated
epoch [18], but not for a string network in the scaling regime.  In
the latter case we need to integrate over all loop sizes to obtain
$(SF)$.

The above analysis does not depend on the topology of the defect in a
key way.  For collapsing domain walls, our mechanism is stronger since
the suppression factor $(SF)$ would be

$$(SF)
\sim
\,
{R_c\over
\xi
(t_{EW})}
\
,
$$

\noindent $R_c$ being proportional to the domain wall core radius.
For collapsing monopoles, however, the mechanism is weaker since

$$(SF) \sim \left( {R_c\over \xi(t_{EW} )} \right)^3\, .$$

There are two ways to increase $(SF)$:  either we decrease $\xi
(t_{EW} )$ or we increase $R_s$.  The obvious way to decrease
$\xi(t_{EW})$ is to decrease the scale $\eta$ of the string producing
phase transition.  The earlier we are in the friction dominated epoch,
the closer the strings are relative to the horizon since [18]

$$\xi (t) \sim \xi (t_f ) \, \left( {t\over t_f }\right)^{5/4} \, ,
\eqno {(4.8)}$$

$t_f$ being the time of string formation (given by $\eta$).  According
to the Kibble mechanism

$$\xi (t_f) \simeq \lambda^{-1}\,\eta^{-1} \, ,\eqno {(4.9)}$$

\noindent where $\lambda$ is the string scalar field self coupling
constant.  As mentioned previously, a cosmic string coupled to the
Weinberg-Salam model restores the electroweak symmetry in a region

$$R_s \sim \eta^{-1}_{EW} \eqno {(4.10)}$$

\noindent and therefore

$$(SF) \sim \left( {\eta_{EW}\over \eta} \right)^3 \, .\eqno {(4.11)}$$

The second way to increase $(SF)$ is to make the strings
superconducting.  For maximal string current this region is

$$R_s \sim {\eta \over \eta^2_{EW}} \eqno {(4.12)}$$

In general, however, the initial current on a superconducting string
will be much less than $\eta$ and thus $R_s$ smaller than (4.12).  To
obtain a more realistic estimate consider the superconductivity to be bosonic
and the winding of the boson condensate at formation to
be $\sim$1.  The winding at later times is $\sim N^{1 \over 2}$, where
$N$ is the ratio of comoving volumes corresponding to loop sizes at
$t$ and $t_f$ respectively.  This winding in the field giving rise to
superconductivity induces a current of order

$$I \sim I_{{\rm max}} (T/\eta)^{1 \over 4} \eqno {(4.13)}$$

\noindent giving

$$R_s \sim (T/\eta)^{1 \over 4} \ {\eta \over \eta^2_{EW}} \eqno
{(4.14)}$$

\noindent and resulting suppression factor of

$$SF \sim \lambda^2 \ ({\eta_{EW} \over \eta})^{3 \over 2} \eqno
{(4.15)}$$

\noindent For $\eta$ just above the electroweak transition and $\lambda$
close to unity the strength of this mechanism for baryogenesis is comparable to
the
first order one.

{\bf \chapter{FUTURE DEVELOPMENTS}}

In the previous section we ignored the effect of the string's
transverse motion and just concentrated on the initial collapse of the
loop.  For superconducting strings this is a gross underestimate.  In
this case the region of symmetry restoration, $R_s$, is significantly
larger than the mean free path of baryons within the string.  As the
string moves, the $CP$ violating phase, $\Delta \theta$, is equal and
opposite on the leading and trailing edges of the string.  At the
leading face anti-baryons are produced.  However, before the trailing
face arrives the anti-baryons have already decayed to anti-leptons and
thus do not annihilate with the baryons produced at the trailing
edge.  The result of this is that the volume of baryogenesis is

$$V_{BG} \sim \xi (t)^2 R_s \left({t \over \xi (t)}\right)^3 \eqno {(5.1)}$$

\noindent rather than (4.4).  The resulting suppression factor becomes

$$SF \sim {R_s \over \xi (t_{EW})} \sim \lambda \left({\eta_{EW} \over
\eta}\right)^{3 \over
4} \eqno {(5.2)}$$

\noindent This makes our mechanism even more attractive.

Finally, we note that the baryogenesis mechanism discussed by Barriola
[19] in the context of electroweak strings can also be applied to
topological defects.  Barriola noted that integrating the anomaly in
the baryon current gives the change in baryon number in a given volume
and time interval

$$\Delta B = {N_f \over 32 \pi^2} \int d^4x \lbrack g^2
E_W.B_W + G^2 \cos(2\theta_W) E_Z.B_Z + {G^2
\over 2} \sin(2 \theta_W) (E_Z.B_A + E_A.B_Z)\rbrack \eqno {(5.3)}$$

\noindent where $E$ and $B$ refer to the electric and magnetic fields
of the physical fields after symmetry breaking, $N_f$ refers to the number of
families and $G^2 = g^2 + g'^2$, where $g$ and $g'$ are the $SU(2)$ and $U(1)$
coupling constants respectively.
In (5.3) the first term is small in the broken phase, the second term
is only non-zero in defects with a non-zero helicity [20], but the
third term could be large.  In the region $R_s$ around a cosmic string
there is a non-trivial $Z$-flux [16].  As discussed at this meeting
[20,21] the electroweak phase transition produces a background
magnetic field of order $B_A \sim gT^2$.  In the two-doublet model
this gives a contribution to the action of

$$ \Delta S = {N_fG^2 \over 64\pi^2} \sin (2\theta_W) \int d^4
x \theta (E_A.B_Z + B_A.E_Z) \eqno {(5.4)}$$

\noindent Integrating by parts this gives a contribution to the free
energy density

$$F_B = -\dot \theta B \eqno {(5.5)}$$

\noindent In the core of collapsing topological defects $\dot \theta > 0$
and so $B$ is driven positive to minimise the free energy.

As a first estimate the rate per unit volume of baryon violation for
electroweak strings is

$$\Gamma_B \sim \gamma_v v_t g^2 T^4 \eqno {(5.6)}$$

\noindent where $v_t$ is the speed at which the loop is collapsing and
$\gamma_v$ the associated Lorentz factor.  This rate is considerably
larger than that resulting from sphaleron processes in (2.1).

The change in baryon number is again given by (4.2), but with rate
$\Gamma_B$ as in (5.6) and volume $V_{BG}$ of (5.1).  Unlike the
mechanism discussed in section 4, this increased rate of baryon
violation allows the exciting possibility of electroweak baryogenesis
with GUT strings.  By the time of the electroweak phase transition the
GUT string network has reached a scaling solution.  Considering the
scaling solution for the number of loops per unit volume and
integrating over loop size, we have estimated the suppression factor
to be $\sim 10^{-1}$ for a scale of symmetry breaking of $10^{15}$ GeV.
This allows the exciting possibility that the same cosmic strings that
may participate in the formation of large scale structure could also
mediate electroweak baryogenesis [22].

To conclude, we have discussed a new mechanism for electroweak
baryogenesis which operates even if the phase transition is second
order.  Topological defects produced in a previous phase transition
can play a role analogous to bubble walls.  We have discussed two
mechanisms for generating an asymmetry of order $n_B/s \sim 10^{-10}$.
The first involves a transition shortly before the electroweak one and
sphaleron processes.  The second, and more speculative, involves a
new, stronger mechanism of baryogenesis and allows the exciting
possibility of GUT strings mediating electroweak baryogenesis.

\medskip

\noindent{\bf ACKNOWLEDGEMENTS}

We wish to thank Tomislav Prokopec for discussions on section 5.  ACD
thanks the organizers of the Sintra meeting, and particularly Filipe
Freire, for inviting her to participate in such a stimulating meeting.

\medskip

\noindent {\bf REFERENCES}

\item {(1.)} N Turok - these proceedings.
\item {(2.)} R Brandenberger and A C Davis - Phys Lett \underbar {B308}
(1993).
\item {(3.)} T Vachaspati - Phys Rev Lett \underbar {68} (1992) 1977.
\item {(4.)} R Brandenberger, A C Davis and M Trodden - Phys Lett \underbar {B}
(in
press).
\item {(5.)} N Manton - Phys Rev \underbar {D28} (1983) 2019, F
Klinkhammer and N Manton - Phys Rev \underbar {D30} (1984) 2212.
\item {(6.)} J F Gunion, H E Haber, G L Kane and S Dawson, `The Higgs
Hunters' Guide' (Addison-Wesley, 1989).
\item {(7.)} N Turok in `Perspectives on Higgs Physics' , ed G Kane
(World Scientific, 1992) and references therein.
\item {(8.)} A Cohen, D Kaplan and A Nelson - Ann Rev Nucl and Part
Sci \underbar {43} (1993) 27 and references therein.
\item {(9.)} M Dine, R Leigh, P Huet, A Linde and D Linde - Phys Rev
\underbar {D46} (1992) 550
\item {(10.)} J Ambjorn, M Laursen and M Shaposhnikov, Phys Lett
\underbar {B197} (1989) 49.
\item {(11.)} H B Nielsen and P Olesen - Nucl Phys \underbar {B61}
(1973) 61.
\item {(12.)} Y Nambu - Nucl Phys \underbar {B130} (1977) 505.
\item {} T Vachaspati - Phys Rev Lett \underbar {68} (1992) 1977.
\item {(13.)} M A Earnshaw and M James - Phys Rev \underbar {D48}
(1993) 5818.
\item {(14.)} M James, L Perivolaropoulos and T Vachaspati - Phys Rev
\underbar {D46} (1992) 5232.
\item {} W B Perkins - Phys Rev \underbar {D47} (1993) 5224.
\item {(15.)} M A Earnshaw and W B Perkins - Phys Let \underbar {B}
(in press).
\item {(16.)} W B Perkins and A C Davis - Nucl Phys \underbar {B406}
(1993) 377.
\item {} M Trodden, BROWN-HET-945 (1994).
\item {(17.)} J Ambjorn and P Olesen - Int J Mod Phys \underbar {A5}
(1990) 4525 and references therein.
\item {(18.)} T W B Kibble - Nucl Phys \underbar {B252} (1985) 227, J
Phys \underbar {A9} (1976) 1387.
\item {(19.)} M Barriola - these proceedings.
\item {(20.)} T Vachaspati - these proceedings.
\item {(21.)} K Enqvist - these proceedings.
\item {(22.)} R Brandenberger, A C Davis, T Prokopec and M Trodden -
in preparation.

\medskip

\noindent{\bf Figure Captions}
\medskip
{\bf Figure 1} A comparison between the electroweak baryogenesis mechanisms
using first order phase transitions (top) and our mechanism (bottom). The
contracting topological defect (bottom) plays a role similar to that of the
expanding bubble wall (top) in that it is the location of extra CP violation
and of baryon number violating processes taking place out of equilibrium.

\end